\documentstyle[aps,pra,preprint]{revtex}

\begin{document}

\title{$SU(n)\otimes U(1)_c$ gauge models with spontaneous symmetry
 breaking}

\author{Ion I. Cot\u aescu}

\address{The West University of Timi\c soara,\\
         V. P\^ arvan Ave. 4, RO-1900 Timi\c soara, Romania}

\date{\today}
\draft
\maketitle

\begin{abstract}
A possible generalization of the technique of the standard model to
$SU(n)\otimes U(1)$ gauge models is proposed.
A special Higgs mechanism and a new kind of Yukawa couplings in unitary
gauge are introduced.
These allow us to obtain a general method of deriving boson mass spectrum
and  coupling coefficients which will be used to find an exact solution
of the Pisano-Pleitez
three-generation $SU(3)\otimes U(1)$ model. A new  anomaly-free
one-generation model is briefly discussed.
\end{abstract}


\section{Introduction}
\

The gauge models with spontaneous symmetry breaking must respect few basic
rules to be renormalizable. The central point is to find a Higgs mechanism
able to break the gauge symmetry up to a desired residual one and to produce
a good boson mass spectrum. Moreover, the
spinor mass terms could arise only from  Yukawa interactions and certainly
the model must be free of the axial anomaly \cite{GENERAL}. In practice these
rules follow to be applied
in each particular case separately since we have not yet a  general
theory of the spontaneous breakdown of an arbitrary unitary gauge symmetry
which satisfies all these requirements. However, if we consider the standard
model
(SM)  and its recent $SU(3)\otimes U(1)$ and $SU(4)\otimes U(1)$
generalizations \cite{PP,FRAM,NG,SU4}, we observe a some kind of regularity,
namely, that the number of
the vector Higgs multiplets may be equal to the dimension of the fundamental
representation. This encourages us to try to find a generalization of
the Higgs mechanisms of these models to any $SU(n)\otimes U(1)$ gauge model.

The problem is complicated since the whole structure of the Higgs sector
depends on the possible Yukawa interactions and, therefore, on the concrete
choice of the representations of the spinor multiplets. In order
to avoid these
 difficulties we shall construct
a special Higgs mechanism and a new kind  of couplings among the spinors
and the Higgs multiplets which will be referred as the {\it minimal Higgs
mechanism} (mHm.). This may help us in the first stage of the model
building when the structure and the parameterization of the model must be
established. After that one can choose if the mHm. will be kept or will be
replaced by an (equivalent or larger) traditional Higgs mechanism.

The mHm. of the $SU(n)\otimes U(1)$ gauge models we would like to propose will
have at  most $n$  Higgs multiplets, each one be of the fundamental
representation of $SU(n)$, but having different chiral hypercharges. In
addition, they will be seen as an orthonormal system
with the "norm" defined by
a scalar real field, which should play the same role as
the neutral component of
the Higgs doublet of the SM.
Here we shall restrict ourselves to present only the simplest
version of this mHm. which will use $n$ such multiplets in order to break the
gauge symmetry up to that of the $U(1)_{em}$ universal gauge. The Lagrangian
density (Ld.) of mHm. will be constructed to have the potential
energy of the simplest form but introducing a metric in its kinetic term to
produce a non-degenerate boson mass spectrum. Furthermore, with
the help of suitable direct products of
Higgs multiplets and with scalar factors
(depending on the "norm" of the vector multiplets), we shall define coupling
terms which
should lead to Yukawa couplings only in unitary gauge. These could produce
adequate mass terms for any pair of spinor components we wish, giving thus more
flexibility to the model. In this article our objective is to derive all the
properties of this class of models with the hope to obtain
a general method to solve  models with high gauge symmetries .

To do this it is convenient to consider  models with the spinor
sector put in the (chiral) pure left (pl.) form, i.e. containing only
left-handed chiral components. This does not restrict the generality since
any model can be brought in this form replacing
all the right-handed components by
the charge conjugated left-handed ones, according to the method largely used in
grand unification theory (GUT) \cite{GENERAL}. For these models  the
group $U(1)$ will coincide with that of the chiral gauge, $U(1)_c$. Therefore
in the following we shall speak about the pl. $SU(n)\otimes U(1)_c$
gauge models.

Starting, in Sec.II, with a short review of the $SU(n)$ representations,
we shall define a suitable hybrid basis of the $su(n)$ algebra and
we shall discuss the $SU(n)\otimes U(1)_c$ representations. In the next one
we shall present the general structure of the spinor and gauge sectors of the
pl. gauge models. The mHm. will be introduced in Sec.IV, where
we shall study the breakdown of the gauge symmetry showing that the residual
symmetry is determined by the hypercharges of the Higgs multiplets.
In addition, we shall define the above mentioned Yukawa coupling
in unitary gauge. The conclusion here is that our mHm. produces the
same effects in unitary gauge as the Higgs mechanism  of the SM.
Based on these results, the properties of the gauge bosons will be
discussed in Sec.V, giving  special attention to the different possible
parameterizations of the model.  There the generalized Weinberg transformation
will be defined and we shall obtain  the masses of the non-hermitian
gauge bosons, the mass matrix of the hermitian ones and
all the coupling coefficients of the charged and neutral currents.
The next section is devoted to the
structure of the spinor sector of the models with  correct electric charge
spectrum,  which should be free of  axial anomaly. This ends with the
summary of the resulted  method  we shall use
 in Sec. VII  to find an exact solution for the Pisano-Pleitez model \cite{PP}
and to discuss a possible new anomaly-free one-generation model.
Finally, we shall briefly comment the role of the mHm. and its possible
extensions.

 \section{Preliminaries}
\

The general treatment of the $SU(n)\otimes U(1)_{c}$ gauge models in  all
their details requires some remarks concerning
the unitary irreducible representations (irep.) of this gauge
group and of its algebra. In this section we shall briefly present their
construction as the direct product between an irep. of $SU(n)$ and an irep.
of $U(1)_{c}$.

\subsection{Conventions and notations}
\

We shall respect the basic notations of the Lie groups and algebras
representation theory \cite{GR,SLAN} with some particularities  which
will be pointed out in the next. Thus the  summation  convention  will  be
systematically
applied only for the pairs of indices in upper and  lower  positions.
This rule
will be extended even to the mathematical structures where these positions are
perfectly equivalent (i.e. the Euclidean spaces and the orthogonal groups).
Moreover, each family of indices we shall use will have its own fixed range,
except the first Greek ones which will be held for the current needs.

Furthermore, we shall simplify the calculations by choosing a unique coupling
constant, $g$, for both the groups involved here. This means that the ireps.
of $U(1)_{c}$, which reduces to multiplications with phase factors
$exp(-igy\xi^{0})$ of parameter $\xi^{0}$, will be defined by the (real)
character $y$ instead of the chiral hypercharge, $y_{ch}$. More precisely,
$y=g'y_{ch}/g$ where $g'$ is the would be specific coupling constant
of $U(1)_{c}$.

The main pieces are the ireps. of $SU(n)$. Their classification can be done
by using the tensor
method \cite{GR}. This starts with the fundamental irep., {\bf n}, which
defines the group $SU(n)$ and its algebra, $su(n)$. If $\xi=\xi^{+}\in su(n)$
then
\begin{equation}\label{(2.1)}
U=U(\xi)=e^{-ig\xi}
\end{equation}
are the unitary matrices of $SU(n)$. These transform the components $\psi_{i}$
of the $n$-dimensional (vector) multiplet, $\psi$, like
\begin{equation}\label{(2.2)}
\psi_{i}\rightarrow(U\psi)_{i}=U_{i\;\cdot}^{\cdot\;j}\psi_{j}
\end{equation}
The transformation law of the complex conjugated components
\begin{equation}\label{(2.3)}
(\psi_{i})^{*}\equiv\psi^{i}\rightarrow(U\psi)_{i}^{*}
\equiv U^{i\;\cdot}_{\cdot\;j}\psi^{j}
\end{equation}
defines the complex conjugated of the fundamental irep., ${\bf n}^*$, which
 has the matrices
$U^{i\;\cdot}_{\cdot\;j}= (U_{i\;\cdot}^{\cdot\;j})^{*}$. All the other ireps.
of $SU(n)$ correspond to the different classes of symmetry (given by the Young
tableaus) of the tensors
of the rank $(r,s)$ provided by the direct products $(\otimes{\bf n})^r\otimes
(\otimes{\bf n}^{*})^s$. These tensors have $r$ lower and $s$ upper
indices for which
we shall reserve the notation, $i,j,k,\cdots=1,\cdots,n$. Their transformation
laws are in accordance with the basic rules given by (\ref{(2.2)})
and (\ref{(2.3)}).

The ireps. of $SU(n)$ will be
denoted either symbolically by $R$ or by indicating their dimension,
$n(R)$, in boldface (as we did already for the ireps. {\bf n} and ${\bf n}^*$).
The notations $R(U)$ and $R(\xi)$ will stand for the transformation matrices
and for the algebra elements of the irep., $R$. The complex conjugated of the
irep. $R$ will be $R^*$ of the matrices $R^*(U)=R(U)^*$.

\subsection{The parameterization of the $su(n)$ algebra}
\

The form of $\xi$ and implicitly that of $R(\xi)$ depends on the
parameterization of the $su(n)$ algebra. The most natural way is to take
the real parameters, $\xi^{a}$, and the hermitian generators,
$T_{a}=T_{a}^{+}$, defined in Appendix A, for which we shall use the indices
$a,b,c,\cdots=1,\cdots,n^2-1$. In this
parameterization $\xi=\xi^{a}T_{a}$. Another possibility is to express:
$\xi=H^{i}_{j}\zeta^{j}_{i}$, by the real generators $H^{i}_{j}$ given by
(\ref{(a.1.1)}) and by the $c$-number parameters
$\zeta^{i}_{j}=(\zeta^{j}_{i})^{*}$.

For the gauge models we intend to work out it is convenient to
introduce a hybrid basis numbered by the indices $i,j,\cdots$ and by a new
family of indices, $\hat i, \hat j,\cdots$, which range from $1$ to $n-1$.
This basis will be defined as follows:
\begin{eqnarray}\label{(2.4)}
D_{\hat i}=T_{(\hat i+1)^{2}-1}\quad,\qquad
E^{i}_{j}=\frac{1}{\sqrt2}H^{i}_{j}
\qquad i\not=j
\end{eqnarray}
The corresponding parameterization:
\begin{equation}
\xi=D_{\hat i}\xi^{\hat i}+\sum_{i\not= j}E^{i}_{j}\xi^{j}_{i}
\end{equation}
contains $n-1$ real parameters, $\xi^{\hat i}$, and $n(n-1)/2$ $c$-number
ones,
$\xi^{i}_{j}=(\xi^{j}_{i})^{*}=\sqrt{2}\zeta^{i}_{j}$, for  $i\not=j$.
This choice
offers many advantages. The first is of a simple numeration of the diagonal
generators, $D_{\hat i}$, (of the Cartan subalgebra). On the other hand, the
parameters $\xi^{i}_{j}$ could be directly associated to the $c$-number gauge
fields because of the factor $1/\sqrt{2}$ from (\ref{(2.4)}) which gives their
correct normalization. However, the most important is to have good trace
orthogonality properties
\begin{eqnarray}\label{(2.6)}
Tr(D_{\hat i}D_{\hat j})=\frac{1}{2}\delta_{\hat i\hat j}\quad &,& \qquad
Tr(D_{\hat i} E^{i}_{j})=0\nonumber\\
Tr(E^{i}_{j}E^{k}_{l})&=&\frac{1}{2}\delta^{i}_{l}\delta^{j}_{k}
\end{eqnarray}
In other respects the commutation relations of this basis
can be calculated according to (\ref{(2.4)}), (\ref{(a.1.2)}) and
(\ref{(a.1.3)}).

\subsection{The irreducible representations of $SU(n)\otimes U(1)_{c}$}
\

Let us denote either by $\rho=(R_{\rho},y_{\rho})$ or by
$\rho=({\bf n}_{\rho},y_{\rho})$,  with $n_{\rho}=n(R_{\rho})$, the irep. of
$SU(n)\otimes U(1)_{c}$ defined as the direct product of the irep. $R_{\rho}$
of $SU(n)$ and the irep. of the character $y_{\rho}$ of $U(1)_{c}$. This will
have the matrices:
\begin{equation}\label{(2.7)}
U^{\rho}(\xi^{0},\xi)=U^{\rho}(\xi)e^{-igy_{\rho}\xi^{0}}=
e^{-ig(\xi^{\rho}+y_{\rho}\xi^{0})}
\end{equation}
where $\xi^{\rho}=R_{\rho}(\xi)$. The transformation law of a
$n_{\rho}$-dimensional multiplet, $\psi^{\rho}$, of this irep. can
be written in the matrix form as
\begin{equation}\label{(2.8)}
\psi^{\rho}\rightarrow U^{\rho}(\xi^{0},\xi)\psi^{\rho}
\end{equation}
understanding that its tensor components (in the direct product basis of the
representation space) transform like:
\begin{equation}\label{(2.9)}
(\psi^{\rho})^{j_{1}\cdots j_{s}}_{i_{1}\cdots i_{r}}\rightarrow
e^{-igy_{\rho}\xi^{0}}U_{i_{1}\;\cdot}^{\;\cdot\;i'_{1}}\cdots
U^{j_{1}\;\cdot}_{\; \cdot\; j'_{1}}\cdots
(\psi^{\rho})^{j'_{1}\cdots j'_{s}}_{i'_{1}\cdots i'_{r}}
\end{equation}
We note that for $y_{\rho}=0$ the irep. $\rho$ coincides with $R_{\rho}$ and
therefore the notations $(R,0)$ and $R$ are equivalent.

The generator of the irep. $\rho$ which corresponds to the $SU(n)$
generator $T$ will be $T^{\rho}=R_{\rho}(T)$. In the next we shall use
the generators defined by (\ref{(2.4)}) and the hermitian ones. Their matrix
elements in the direct product  basis can be calculated according to the
tensor method. For example, if $D$ is a diagonal $SU(n)$ generator, of
matrix elements $D_{i}^{j}=d_{i}\delta^{j}_{i}$, then $D^{\rho}$ will be also
diagonal and we have:
\begin{equation}\label{(2.10)}
(D^{\rho}\psi^{\rho})^{j_{1}\cdots j_{s}}_{i_{1}\cdots i_{r}}=
\left(\sum_{\alpha=1}^{r} d_{i_{\alpha}}- \sum_{\beta=1}^{s}
d_{j_{\beta}}\right)
(\psi^{\rho})^{j_{1}\cdots j_{s}}_{i_{1}\cdots i_{r}}
\end{equation}
It is clear that the $U(1)_{c}$ generator of the irep. $\rho$ is
$T_{0}^{\rho}=y_{\rho}1_{\rho}$ where $1_{\rho}$ is the unit matrix of the
irep. $R_{\rho}$ (to be omitted when no confusion danger is present).

The complex conjugated of the irep. $\rho$ will be
$\rho^{*}=(R_{\rho}^{*},-y_{\rho})$.
This transforms the tensor components of $(\psi^{\rho})^{*}$ which are related
to those of $\psi^{\rho}$ according to the raising an lowering rule of
indices defined by (\ref{(2.3)}).
The hermitian $SU(n)$ generators of the irep. $\rho^{*}$ are
\begin{equation}\label{(2.12)}
T_{a}^{\rho^{*}}=-(T_{a}^{\rho})^{T}
\end{equation}
while those defined by (\ref{(2.4)}) become
\begin{equation}\label{(2.11)}
D^{\rho^{*}}_{\hat i}=-D^{\rho}_{\hat i}\quad,\qquad
E^{\rho^{*}\;i}_{\;\;\;\;j}=-E^{\rho\; j}_{\;\;\; i}
\end{equation}
if the parameterization remains unchanged.
The complex conjugation changes also the sign of the $U(1)_{c}$ generator.

\section{The pure left gauge models}
\

The Ld. of any gauge model with spontaneous symmetry breaking has three terms
\begin{equation}
{\cal L}={\cal L}_{S}+{\cal L}_{G}+{\cal L}_{H}
\end{equation}
They correspond to the spinor (${\cal L}_{S}$), gauge (${\cal L}_{G}$) and
Higgs (${\cal L}_{H}$) sectors \cite{GENERAL}. Let us start with
an arbitrary spinor sector.

\subsection{The pure left form of the spinor sector}
\

In general any spinor sector contains  a given number of left and right-handed
chiral components. After second quantization all the spinor fields will be
fermions. Thus it is natural to suppose from the beginning that all
their components anticommute among themselves (as Grassmann classical
variables or as quantum fermion fields). Thereby we have the possibility to
replace each right-handed component by a charge conjugated left-handed one, as
it is shown in Appendix B. BY doing this we shall obtain a pl. model
containing  only the left-handed components grouped in the multiplet
$L=\left( L_{1},L_{2},\cdots,L_{N}\right)^{T}$ the Dirac adjoint of which is
$\overline L=\left(\overline L_{1},\overline L_{2},\cdots,
\overline L_{N}\right)$. (The superscript $T$ stands for matrix
transposition.) The components of these two multiplets represent
a set of $2N$ independent Grassmann formal variables (each one being in fact
a 2-component spinor). Thus the spinor sector of any model can be put
in the  pl. form. The number $N$ of the left-handed components will
give the dimension of the model.

The most general form of the Ld. of the spinor sector of a pl.
model can be written in the compact matrix notation as:
\begin{equation}\label{(3.2)}
{\cal L}_{S_{0}}=\frac{i}{2}\overline L\tensor{\not\partial} L-
          \frac{1}{2}(\overline L \chi L^{c}+\overline{L^{c}}\chi^{+}L)
\end{equation}
where $\tensor{\not\partial}=\gamma^{\mu}\roarrow{\partial_{\mu}}-
\gamma^{\mu}\loarrow{\partial_{\mu}}$ and
$L^{c}=\left(L_{1}^{c},L_{2}^{c},\cdots,L_{N}^{c}\right)^{T}$ is the charge
conjugated of the multiplet $L$.
The field $\chi$ has been introduced to give rise to the spinor masses after
the breakdown of the gauge symmetry and therefore this may depend on the
Higgs field components. Moreover, according to (\ref{(a.2.5)}), $\chi$ results
to be a $N\times N$ symmetric matrix of scalar fields which could
couple all the spinor components among themselves, being able to generate
simultaneously the Dirac  and Majorana mass terms (like those of
(\ref{(a.2.7)}) and (\ref{(a.2.9)}) respectively), in the limits of the
possible Yukawa interactions.

We observe that generally in (\ref{(3.2)})
quadratic terms in the anticommuting spinor variables could arise.
Consequently,
the correct Euler-Lagrange equations will be obtained by using  Grassmann
derivatives.

\subsection{The global symmetries}
\

One of the advantages of the pl. form of the spinor sector is to point
out the maximal global symmetry of the model, which is partially hidden when
the left and right-handed multiplets are treated separately. This will be
defined as the global symmetry of the kinetic part (i.e. the first term) of
(\ref{(3.2)}). It is clear that the maximal symmetry is given by the group
$SU(N)\otimes U(1)_c$
since the kinetic term remains invariant if the multiplets
$L$ and $\overline L$ transform according to the ireps. $({ \bf N}, y)$
and $({ \bf N}^{*}, -y)$ of this group, for any $y$.
Moreover, if $\chi$ would behave as a symmetric tensor of the irep.
$\left({ \bf N(N+1)/2}, 2y\right)$ of the same group then the whole Ld.
of the spinor sector will have this symmetry. Hence, as long as we do not
make supplementary restrictive hypotheses
relative to the covariance properties of
the field $\chi$, the maximal global symmetry of the spinor sector is
determined only by its dimension, $N$. Its knowledge is important for the
choice of the gauge group since this can be any subgroup of
$SU(N)\otimes U(1)_{c}$.

Let us choose the group $SU(n)\otimes U(1)_{c}$, with $n<N$, to be the gauge
group of our pl. model. Then $L$ will be of a reducible
representation defined as the direct sum of a given set
of ireps., $\rho$. Therefore, we have,
\begin{equation}
L=\sum_{\rho}\oplus L^{\rho}
\end{equation}
where each multiplet, $L^{\rho}$, transforms according to its own irep.,
$\rho$, defined by (\ref{(2.9)}). The corresponding charge conjugated
multiplets, $(L^{\rho})^c$, will transform according to the complex conjugated
ireps., $\rho^{*}$. Furthermore, we see that the matrix $\chi$ contains all
the blocks, $\chi^{\rho\rho'}$, which couple the pairs of the multiplets
$\overline L^{\rho}$ and $(L^{\rho'})^{c}$. Then the Ld. (\ref{(3.2)}) can
be written as:
\begin{equation}\label{(3.3)}
{\cal L}_{S_{0}}=\frac{i}{2}\sum_{\rho}\overline {L^{\rho}}
        \tensor{\not\partial} L^{\rho}-
          \frac{1}{2}\sum_{\rho\rho'}\left(\overline {L^{\rho}}
\chi^{\rho\rho'}(L^{\rho'})^{c}+h.c.\right)
\end{equation}
This remains invariant under the global $SU(n)\otimes U(1)_{c}$
transformations
if the blocks $\chi^{\rho\rho'}$ will transform like
\begin{equation}\label{(3.4)}
\chi^{\rho\rho'}\rightarrow U^{\rho}(\xi^{0},\xi)\chi^{\rho\rho'}
                \left(U^{\rho'}(\xi^{0},\xi)\right)^{T}
\end{equation}
according to the representations $(R_{\rho}\otimes R_{\rho'},y_{\rho}+
y_{\rho'})$ which generally are reducible.

We note that there some discrete symmetries could be added. These  can be
chosen from the list of the discrete groups of the coset space $SU(N)/SU(n)$.

\subsection{Gauging $SU(n)\otimes U(1)_c$}
\

Now we shall gauge the $SU(n)\otimes U(1)_{c}$ group and, consequently, we
shall introduce the gauge fields: $A^{0}_{\mu}=(A^{0}_{\mu})^{*}$ and
$A_{\mu}=A_{\mu}^{+}\in su(n)$ which
can be expressed in the basis (\ref{(2.4)})
as:
\begin{equation}\label{(3.5)}
A_{\mu}=D_{\hat i}A^{\hat i}_{\mu}+\sum_{i\not= j}E^{i}_{j}A^{j}_{i\mu}
\end{equation}
This depends on $n-1$ real fields, $A^{\hat i}_{\mu}$, and $n(n-1)/2$
$c$-number ones which satisfy: $A^{j}_{i\mu}=(A^{i}_{j\mu})^{*}, i\not= j$.
The next step is to replace the ordinary derivatives of the Ld (\ref{(3.3)})
by the covariant ones:
\begin{equation}
D_{\mu}L^{\rho}=\partial_{\mu}L^{\rho}-ig(R_{\rho}(A_{\mu})+
                       y_{\rho}A_{\mu}^{0})L^{\rho}
\end{equation}
which will give the interaction terms of the whole Ld. of the spinor sector.
In the basis (\ref{(2.4)}) this is:
\begin{equation}\label{(3.8)}
{\cal L}_{S}={\cal L}_{S_{0}}+g \sum_{\rho}\overline {L^{\rho}}
           \left(D^{\rho}_{\hat i}A^{\hat i}_{\mu}+
                  \sum_{i\not= j}E^{\rho\; i}_{\;\;\;j}A^{j}_{i\mu}+
y_{\rho}A^{0}_{\mu}\right)\gamma^{\mu}L^{\rho}
\end{equation}

The gauge invariance of this Ld. requires the gauge fields to transform like:
\begin{eqnarray}
A_{\mu}&\rightarrow& UA_{\mu}U^{+}-\frac{i}{g}(\partial_{\mu}U)U^{+}\\
A_{\mu}^{0}&\rightarrow& A_{\mu}^{0}+\partial_{\mu}\xi^{0}\nonumber
\end{eqnarray}
where $U=U(\xi(x))$ is given by (\ref{(2.1)}). The field strength tensors:
\begin{eqnarray}\label{(3.10)}
F_{\mu\nu}&=&\partial_{\mu}A_{\nu}-\partial_{\nu}A_{\mu}
                           -ig\left[A_{\mu},A_{\nu}\right]\\
F_{\mu\nu}^{0}&=&\partial_{\mu} A_{\nu}^{0}-\partial_{\nu}A_{\mu}^{0}\nonumber
\end{eqnarray}
are covariant and consequently we can define the invariant Ld. of
the gauge sector as
\begin{equation}\label{(3.11)}
{\cal L}_{G}=-\frac{1}{2}Tr\left(F_{\mu\nu}F^{\mu\nu}\right)-
                  \frac{1}{4} F_{\mu\nu}^{0}F^{0\mu\nu}
\end{equation}

All these relations can be written in the
basis (\ref{(2.4)}) starting with the form of the matrix $F_{\mu\nu}$ in
this basis:
\begin{equation}
F_{\mu\nu}=D_{\hat i}F^{\hat i}_{\mu\nu}+
\sum_{i\not= j}E^{i}_{j}F^{j}_{i\mu\nu}
\end{equation}
where, according to (\ref{(2.6)}), we have
\begin{equation}
F_{\mu\nu}^{\hat i}=2Tr(D_{\hat i}F_{\mu\nu})\quad,\quad
        F^{j}_{i\mu\nu}=(F^{i}_{j\mu\nu})^{*}=2Tr(E^{j}_{i}F_{\mu\nu})
\end{equation}
Furthermore, the Ld. (\ref{(3.11)}) will get the form:
\begin{equation}\label{(3.14)}
{\cal L}_{G}=-\frac{1}{4}\left(\sum_{\hat i}F_{\mu\nu}^{\hat i}F^{\hat i\mu\nu}
                   +F_{\mu\nu}^{0}F^{0\mu\nu}\right)-
               \frac{1}{2}\sum_{i}\sum_{j<i}\left(F_{i\mu\nu}^{j}
\right)^{*}F_{i}^{j\mu\nu}
\end{equation}
Its kinetic part, ${\cal L}_{G\rfloor g=0}$, will contain only the
kinetic parts of the  fields, $F_{\mu\nu}^{0}$ and $F_{\mu\nu\rfloor
g=0}$. This will be combined with the mass term resulted from the breakdown of
the gauge symmetry, in order to obtain the free Ld. of the gauge sector.

\section{The minimal Higgs mechanism}
\

Our aim is to introduce the mHm. as a generalization the Higgs mechanism
 of the SM by using a minimal number of field
variables and a simple Ld. which should take in unitary gauge the same
familiar form as that of SM. Moreover, this mHm. must  break the
$SU(n)\otimes U(1)_{c}$ gauge symmetry up to that of the $U(1)_{em}$ producing
only one nonvanishing vacuum expectation value (vev.). This means that only
one scalar real field, $\phi$, should survive  the gauge fixing.
On the other hand, it is clear that in these conditions
the blocks of $\chi$, which generally could be tensors of any
rank, does not coincide with the Higgs fields even though they may have
nonvanishing vevs..
The solution is to suppose that the $\chi$-blocks are direct products of Higgs
multiplets, in accordance to their gauge covariance, but having suitable
 scalar factors
which should lead to the Yukawa couplings only in unitary gauge. However,
these factors may depend only on the scalar $\phi$.

\subsection{The Higgs sector}
\

Let us take $n^{2}$ $c$-number Higgs variables, $\phi^{(i)}_{j}$,
organized into $n$ multiplets, $\phi^{(i)}$, each one transforming according
its own irep., $({\bf n},y^{(i)})$. These represent a set of $2n^2$ real field
variables from which at most $n^2$ could be removed by fixing the gauge.
Thus the danger is to remain with some Goldstone bosons after the breakdown of
the gauge symmetry. To avoid this, we shall reduce the number of field
variables by introducing a priori the following {\it constraints}:
\begin{equation}\label{(4.1)}
{\phi^{(i)}}^{+}\phi^{(j)}=\phi^{2}\delta_{ij}
\end{equation}
where $\phi$ is a gauge invariant real field variable. These are $n^2$ real
equations and, therefore, the fields $\phi^{(i)}_j$ will have only $n^2$
independent real components.  We note that the indices
included in parentheses are not $SU(n)$ vectorial ones even though they still
range from $1$ to $n$. Thus, in fact, we have introduced an orthonormal basis
in the representation space of the irep.
{\bf n} of $SU(n)$ the vectors of which
transform differently under the $U(1)_{c}$ group. Their $U(1)_{c}$
characters, $y^{(i)}$, will be considered as arbitrary parameters. They can be
grouped into the matrix:
\begin{equation}\label{(4.2)}
Y=diag\left(y^{(1)},y^{(2)},\cdots,y^{(n)}\right)
\end{equation}

In order to obtain a Higgs Ld. with very simple interaction terms but able to
produce a non trivial boson mass spectrum,
it needs to introduce free parameters
in its kinetic term. These will be: $\eta_{0}\in[0,1)$ and the nonvanishing
elements $\eta^{(i)}$ of the matrix $\eta=diag\left(\eta^{(1)},\eta^{(2)},
\cdots,\eta^{(n)}\right)$ with the property,
\begin{equation}\label{(4.3)}
Tr(\eta^{2})=1-{\eta_{0}}^{2}
\end{equation}
Now we can use $({\eta_{0}}^{2}, \eta^2)$ as the metric of the kinetic part of
the Ld. of the Higgs sector which will be
\begin{equation}\label{(4.4)}
{\cal L}_{H}=
\frac{1}{2}{\eta_{0}}^{2}\partial^{\mu}\phi \partial_{\mu}\phi +
\frac{1}{2}\sum_{i}(\eta^{(i)})^{2}
\left({\cal D}^{\mu}\phi^{(i)}\right)^{+}\left({\cal D}_{\mu}\phi^{(i)}\right)
              -V(\phi)
\end{equation}
where
\begin{equation}\label{(4.5)}
{\cal D}^{\mu}\phi^{(i)}=\partial_{\mu}\phi^{(i)}-
              ig\left(A_{\mu}+y^{(i)}A_{\mu}^{0}\right)\phi^{(i)}
\end{equation}
and
\begin{equation}\label{(4.6)}
V(\phi)=-\frac{1}{2}\mu^2\sum_{i}{\phi^{(i)}}^+\phi^{(i)}+
   \frac{1}{4}\lambda_{1}\left(\sum_{i}{\phi^{(i)}}^{+}\phi^{(i)}\right)^2+
   \frac{1}{4}\lambda_{2}\sum_{i,j}\left({\phi^{(i)}}^{+}\phi^{(j)}\right)
               \left({\phi^{(j)}}^{+}\phi^{(i)}\right)
\end{equation}
is the potential which has been chosen to have the simplest algebraic form
(allowed by its gauge invariance).

Furthermore, we shall look for the absolute minimum of $V(\phi)$. To this end,
we can apply two methods. The first one is to give up
(\ref{(4.1)}) and to require the vanishing of all the derivatives of
$V(\phi)$ with respect to $\phi^{(i)}_{j}$. This leads to the equations:
\begin{equation}\label{(4.7)}
\delta_{ij}\left(-\mu^{2}+\lambda_{1}\sum_{k}{\phi^{(k)}}^{+}\phi^{(k)}\right)+
                \lambda_{2}{\phi^{(i)}}^{+}\phi^{(j)}=0
\end{equation}
which have as solutions only  the set of orthonormal multiplets (\ref{(4.1)})
with $\phi$
given by:
\begin{equation}\label{(4.8)}
-\mu^{2}+(n\lambda_{1}+\lambda_{2})\phi^{2}=0
\end{equation}
Another possibility is to express (\ref{(4.6)}) only on $\phi$ with the help
of (\ref{(4.1)}) and to take the minimum in this unique variable.
The result is the same, namely (\ref{(4.8)}). Thus it results that the
constraints (\ref{(4.1)}) are compatible with the (absolute) minimum
of $V(\phi)$. This will define the vacuum state in which $\phi$ will have a
nonvanishing expectation value, $<\phi>$. Then, $\phi=<\phi>+\sigma$
where $\sigma$ is the physical Higgs field (with zero vev.). From
(\ref{(4.8)}) we obtain in the zero-th order of the perturbations
\begin{equation}\label{(4.22)}
 <\phi>=\mid\mu\mid/\sqrt{n\lambda_{1}+\lambda_{2}}
\end{equation}
Moreover, because of (\ref{(4.1)}),  we are sure that here exists a gauge in
which
\begin{equation}\label{(4.11)}
\hat \phi^{(i)}_{k}=\delta_{ik}\phi=\delta_{ik}(<\phi>+\sigma)
\end{equation}
This will be the unitary gauge. To go now to an arbitrary gauge one needs to
perform a ("boost") transformation, $\hat U=\hat U(\phi^{(i)})$, so that
$\phi^{(i)}=\hat U \hat\phi^{(i)}$.
Therefore, the components of $\phi^{(i)}$ in an arbitrary gauge can be written
as:
\begin{equation}\label{(4.21)}
\phi^{(i)}_{j}=\hat U_{j\;\cdot}^{\cdot\;i}(<\phi>+\sigma)
\end{equation}

The Ld. of the Higgs sector in unitary gauge can be calculated by using
(\ref{(4.3)})-(\ref{(4.6)}), and (\ref{(4.11)}). We find  that this is
\begin{eqnarray}\label{(4.15)}
{\cal L}_{H}=\frac{1}{2}\partial^{\mu}\sigma\partial_{\mu}\sigma &-&
             \frac{1}{2}m_{\sigma}^{2}\sigma^{2}-
             \frac{m_{\sigma}^{2}}{2<\phi>}\sigma^{3}-
             \frac{m_{\sigma}^{2}}{8<\phi>^{2}}\sigma^{4}+\nonumber\\
 &+&\frac{g^{2}}{2}\left(<\phi>+\sigma\right)^{2}Tr[(A_{\mu}+YA_{\mu}^{0})
           \eta^{2}(A^{\mu}+YA^{0\mu})]
\end{eqnarray}
where $m_{\sigma}=\sqrt{2n}\mid\mu\mid$ is the mass of the field $\sigma$.
We see that it has the same form like in the SM (or in the Weinberg- Salam
model \cite{WS,KH}), except the last term, which  contains here the both kinds
of the free parameters introduced above ($y^{(i)}$ and $\eta^{(i)}$). This will
produce the boson mass spectrum which will be analyzed in more detail in
the next section.

Particularly, for $n=2$, the pair of the Higgs doublets of the SM, $\phi_{SM}$
and $\tilde\phi_{SM}$ \cite{KH}, can be recognized to be in our notation:
\begin{equation}\label{(4.16a)}
\frac{1}{\sqrt{2}}\phi^{(2)}=\phi_{SM}=
\begin{array}{|c|}
\phi_+\\
\phi_0
\end{array} \quad;\quad
\frac{1}{\sqrt{2}}\phi^{(1)}=\tilde\phi_{SM}=
\begin{array}{|c|}
{\phi_0}^*\\
-{\phi_+}^*
\end{array}
\end{equation}
Moreover, it can be proved that in this case the Ld. (\ref{(4.15)}) with
$\eta_{0}=0$ and the homogeneous metric,
\begin{equation}\label{(4.17a)}
\eta^{(1)}=\eta^{(2)}=1/\sqrt{2}
\end{equation}
coincides with that of the SM. Thus, the mHm. appears as a natural
generalization of the Higgs mechanism of the SM.

\subsection{The residual symmetry}
\

{}From the apparently large collection of Higgs field variables, we remain
only with the physical scalar $\sigma$. However, the other ones (i.e. the
Goldstone bosons) do
not disappear without trace, since they provide us with the
characters $y^{(i)}$ which will be involved in all the sectors of the model.

These will determine even the existence of the residual
symmetry. This is given by the little group which leaves invariant the form
(\ref{(4.11)}) of the Higgs fields in  unitary gauge. Therefore, its
transformations must satisfy
\begin{equation}
e^{-ig(\xi+y^{(i)}\xi^{0})}\hat \phi^{(i)}=\hat \phi^{(i)}
\end{equation}
for all $i=1\cdots n$. This requires $\xi$ to be from the
Cartan subalgebra, i.e. $\xi=D_{\hat i}\xi^{\hat i}$, and to have:
\begin{equation}\label{(4.13)}
D_{\hat i}\xi^{\hat i}+Y\xi^{0}=0
\end{equation}
The nontrivial solutions of this homogeneous system will span
the parameter space of the little group. These could arise only if $Y$
will be also of the Cartan subalgebra so that:
\begin{equation}\label{(4.14)}
Tr(Y)=0
\end{equation}
Then the system (\ref{(4.13)}) will admit as a unique nontrivial solution the
one-dimensional subspace of the equations
$\xi^{\hat i}+2\xi^{0}Tr(D_{\hat i}Y)=0$ and nothing else. This will be
just the subspace of the parameter $\xi^{em}$ of the little group $U(1)_{em}$.
In this case $\hat U$ will be a transformation of the coset space
$SU(n)\otimes U(1)_{c}/U(1)_{em}$, (depending on $n^{2}-1$ real parameters).
Moreover,  the condition (\ref{(4.14)}) will reduce to $n-1$ the number of the
independent characters, $y^{(i)}$. Their concrete values will
determine the direction of $\xi^{em}$ in the $n$-dimensional parameter subspace
$\lbrace\xi^{0},\xi^{\hat i}\rbrace$.

Hence, it is obvious that the mHm. with $n$ vector multiplets is able to break
the $SU(n)\otimes U(1)_c$
gauge symmetry up to the $U(1)_{em}$ residual one, when the condition
(\ref{(4.14)}) is accomplished. If it will be applied to the
$SU(n)\otimes U(1)_c$ subgroup  a $S(n')\otimes SU(n)\otimes U(1)_c$  gauge
group then the residual symmetry group will be $SU(n')\otimes U(1)_{em}$. This
means that this mHm. could be used  for the actual generalizations of the SM.
We note that the mHm. with a reduced number of vector multiplets will lead to
larger residual symmetries. Thus if we shall use only $n-k$ orthogonal
multiplets in a $SU(n)\otimes U(1)_c$ gauge model then a $SU(n-k)$ symmetry
will be broken and the residual symmetry will be given by $SU(k)\otimes U(1)$
(where $SU(k)$ is the maximal subgroup of the coset space $SU(n)/SU(n-k)$).

\subsection{The Yukawa couplings in unitary gauge}
\

As we have mentioned, to be in accordance with the gauge
covariance, we must define each block of $\chi$ as a direct product of some
Higgs multiplets. In addition we shall introduce factors of the form
$\phi^{-p}$ to obtain Yukawa couplings in unitary gauge.

Let us take the block $\chi^{\rho\rho'}$
which transforms like (\ref{(3.4)}) according to the irep.
$\left(R_{\rho}\otimes R_{\rho'},y_{\rho}+y_{\rho'}\right)$. If the ireps.
$\rho$ and $\rho'$ are the ranks $(r,s)$ and $(r',s')$ respectively, then we
find the components of $\chi^{\rho\rho'}$ to be tensors of the rank
$(r+r',s+s')$. These will be defined as follows:
\begin{eqnarray}\label{(4.16)}
\left(\chi^{\rho\rho'}\right)^
{j_{1}\cdots j_{s},j'_{1}\cdots j'_{s'}}_{i_{1}\cdots i_{r},i'_{1}
\cdots i'_{r'}}=\phi^{-p}\sum_{k,l}
       G^{l_{1},l_{2},\cdots,l_{s+s'}}_{k_{1},k_{2},\cdots,k_{r+r'}}\times
    \phi^{(k_{1})}_{i_{1}}\cdots\phi^{(k_{p+p'})}_{i'r'}\times
    \left(\phi^{(l_{1})}_{j_{1}}\right)^{*}\cdots
    \left(\phi^{(l_{q+q'})}_{j's'}\right)^{*}
\end{eqnarray}
where $G^{l_{1}\cdots}_{k_{1}\cdots}$ are coupling constants. This form
corresponds to the reducible representation
$R_{\rho}\otimes R_{\rho'}$ of $SU(n)$ but
to fix the value of its character to
$y_{\rho}+y_{\rho'}$ it requires a supplementary selection rule. This is:
the coupling constants of (\ref{(4.16)}) can have non-zero arbitrary values
only for those combinations of indices for which we have:
\begin{equation}\label{(4.17)}
y^{(k_{1})}+\cdots +y^{(k_{p+p'})}-\left( y^{(l_{1})}+\cdots +y^{(l_{q+q'})}
                                    \right)=y_{\rho}+y_{\rho'}
\end{equation}
When this condition is not satisfied the coupling constants must vanish.
Thus the components of $\chi$ are well defined according to the gauge
invariance of ${\cal L}_{s}$.

In (\ref{(4.16)}) we have introduced the scalar factor,
$\phi^{-p}$, to control the formal dimensions of the coupling terms
\cite{IZ,COL}. It is known that the model will be renormalizable only
if each block $\chi^{\rho\rho'}$ will be of
the  dimension $d(\chi^{\rho\rho'})\le 1$. These can be easily pointed out
in unitary gauge where (\ref{(4.16)}) becomes:
\begin{equation}\label{(4.18)}
\left(\hat \chi^{\rho\rho'}\right)^
{j_{1}\cdots j_{s},j'_{1}\cdots j'_{s'}}_{i_{1}\cdots i_{r},i'_{1}
\cdots i'_{r'}}=\left(<\phi>+\sigma\right)^{d(\chi^{\rho\rho'})}
       G^{j_{1}\cdots j_{s},j'_{1}\cdots j'_{s'}}_
        {i_{1}\cdots i_{r},i'_{1}\cdots i'_{r'}}
\end{equation}
with the obvious identification
\begin{equation}
d(\chi^{\rho\rho'})=r+r'+s+s'-p
\end{equation}
Thus the power $p$ of the scalar factor will be the parameter giving
the desired formal dimension of the coupling terms. This will be fixed to
$p=r+r'+s+s'-1$ in order to obtain the Yukawa couplings in unitary gauge (when
$d(\chi^{\rho\rho'})=1$). In the other possible case (of
$d(\chi^{\rho\rho'})=0$) the field $\phi$ is completely decoupled and the mass
terms appear as put by hand.

Here it is important to note that the factor $\phi^{-p}$ can not produce
singularities as long as $p$ is at  most equal to the number of the fields
$\phi^{(i)}$ of (\ref{(4.16)}). The argument is that $\phi^{(i)}_{j}/\phi$
results from (\ref{(4.21)}) to be just a matrix element of the "boost"
$\hat U$ which must be regular. On the other hand, we observe that, despite
the unusual definition of the $\chi$-blocks, their form in unitary gauge
leads to the familiar couplings among  $\sigma$ and spinors like in
the SM. Thus we can conclude that the mHm. produces
the same effects in unitary gauge as the Higgs mechanism of the SM.
In addition it has  the advantage to be able to couple any pair of spinor
multiplets in the mass terms, with the unique restriction given by
(\ref{(4.17)}).

\section {The physical gauge bosons and their coupling coefficients}
\

Generally, the $c$-number gauge fields can be directly
associated to the non-hermitian gauge bosons but the real gauge fields are
linear combinations of the hermitian physical fields. These arise from some
global transformations which leave invariant the kinetic part of the Ld.
(\ref{(3.14)}) and diagonalize the mass matrix generated by the last term of
(\ref{(4.15)}). In the following we shall explicitly obtain these
transformation which will help us to calculate the coupling coefficients of
the charged and neutral currents i.e., the electric and the neutral charges.

\subsection{The separation of the electromagnetic potential}
\

The first step is to extract the electromagnetic potential, $A_{\mu}^{em}$,
associated to the $U(1)_{em}$ parameter, $\xi^{em}$. In the previous section
we have seen that $\xi^{em}$ belongs to a direction, of the parameter subspace
$\{\xi^{0},\xi^{\hat i}\}$, depending on the values of the characters
(\ref{(4.2)}) which  satisfy the condition (\ref{(4.14)}). We shall try to
separate $A_{\mu}^{em}$ from the other real gauge fields by changing the basis
of $\{\xi^{0},\xi^{\hat i}\}$ in order to bring $\xi^{em}$ along  the  new
zero-direction.

This can be done since the kinetic part of the first term  of the Ld
(\ref{(3.14)}) is invariant under the $SO(n)$ global transformations of the
subspace $\{\xi^{0},\xi^{\hat i}\}$ of  the form:
\begin{eqnarray}\label{(5.1)}
\xi^{0}&=&V^{0\; \cdot}_{\; \cdot \; 0}\xi'^{0}+
        V^{0\; \cdot}_{\; \cdot \; \hat i}\xi'^{\hat i}\nonumber\\
\xi^{\hat i}&=&V^{\hat i\; \cdot}_{\; \cdot \; 0}\xi'^{0}+
       V^{\hat i\; \cdot}_{\; \cdot \; \hat j}\xi'^{\hat j}
\end{eqnarray}
which change like the group parameters, the gauge fields as well as the kinetic
parts of the field strength tensors. We shall choose a special
$SO(n)$ transformation defined as a rotation of  angle $\theta$ around
the axis of versor $\nu$ orthogonal to the $\xi^{0}$ direction
(i.e. $\nu_{0}=0$ , $\nu_{\hat i}=\nu^{\hat i}$ and $\nu_{\hat i}\nu^{\hat i}
=1$) and we shall require $\xi'^{0}$ to coincide with $\xi^{em}$. This
transformation has the matrix elements:
\begin{eqnarray}\label{(5.2)}
V^{0\; \cdot}_{\; \cdot \; 0}&=&\cos\theta\nonumber\\
V^{\hat i\; \cdot}_{\; \cdot \; 0} &=&-V^{0\; \cdot}_{\; \cdot \; \hat i}=
\nu_{\hat i}\sin\theta\\
V^{\hat i\; \cdot}_{\; \cdot \;\hat j}&=&\delta^{\hat i}_{\hat j}-
\nu^{\hat i}\nu_{\hat j}(1-\cos\theta)\nonumber
\end{eqnarray}
Now we see that $\xi^{em}$ will be the unique solution of (\ref{(4.13)})
only if we have:
\begin{equation}\label{(5.3)}
Y=-D_{\hat i}\nu^{\hat i}\tan\theta \equiv -(D\cdot\nu)\tan\theta
\end{equation}
With the help of (\ref{(2.6)}),  we can calculate the values of
$\theta$ and $\nu_{\hat i}$ for a given matrix $Y$ which satisfy the
condition (\ref{(4.14)}). Thus the transformation
(\ref{(5.2)}) leading the gauge fields $A^{0}_{\mu}$ and $A^{\hat i}_{\mu}$
into the new ones, $A^{em}_{\mu}$ and $A'^{\hat i}_{\mu}$, is completely
determined.

On the other hand, (\ref{(5.3)}) can be interpreted as the change of the
$n-1$ arbitrary parameters, $y^{(i)}$, to the new ones, $\theta$ and
$\nu_{\hat i}$ (with $n-2$ independent components). In the next
we shall use this new parameterization which will be  more efficient.

\subsection{The massive gauge bosons}
\

Let us take the mass term of the gauge fields given by the last term of the
Ld. (\ref{(4.15)}). This is:
\begin{equation}
\frac{g^{2}}{2}<\phi>^{2}Tr \left[\left(A_{\mu}+YA^{0}_{\mu}\right)\eta^{2}
                                  \left(A^{\mu}+YA^{0\mu}\right)\right]
\end{equation}
where $\eta^{2}$ satisfies the condition
(\ref{(4.3)}). It can be expressed in terms of the fields
$A^{em}_{\mu}$ and $A'^{\hat i}_{\mu}$ and of the parameters $\theta$ and
$\nu_{\hat i}$ instead of $y^{(i)}$. Thus by
(\ref{(3.5)}), (\ref{(5.2)}) and (\ref{(5.3)}), after few manipulations, we
can put it in the form:
\begin{equation}
\frac{1}{2}(M^{2})_{\hat i \hat j} {A'_{\mu}}^{\hat i} {A'}^{\hat j \mu}+
     \sum_{i} \sum_{j<i} \left ( M^{j}_{i}\right )^{2}
\left (A^{j}_{i\mu}\right)^{*}
       A^{j\mu}_{i}
\end{equation}
Here $M^{2}$ is a non-diagonal matrix with the elements given by
\begin{equation}\label{(5.8)}
(M^{2})_{\hat i\hat j}=<\phi>^{2}Tr(B_{\hat i}B_{\hat j})
\end{equation}
where
\begin{equation}\label{(5.23)}
B_{\hat i}=g\left(D_{\hat i}+\nu_{\hat i}(D\cdot \nu)
                       \frac{1-\cos\theta}{\cos\theta}\right)\eta
\end{equation}
while
\begin{equation}\label{(5.22)}
M^{j}_{i}=\frac{1}{2}g<\phi>\left((\eta^{(i)})^{2}+
(\eta^{(j)})^{2}\right)^{1/2}
\end{equation}
are just the masses of the $c$-number gauge fields, $A^{j}_{i\mu}$.

As it was expected, $A^{em}_{\mu}$ does not appear in the mass term and,
consequently, it remains massless. The other real gauge fields
${A'}_{\mu}^{\hat i}$ have the non-diagonal mass matrix (\ref{(5.8)}). This
can be brought in
diagonal form with the help of a new $SO(n-1)$ transformation:
\begin{equation}\label{(5.11)}
A'^{\hat i}_{\mu}=\omega^{\hat i\;\cdot}_{\cdot\;\hat j}Z_{\mu}^{\hat j}
\end{equation}
which will lead to the neutral gauge bosons $Z_{\mu}^{\hat i}$ with
well-defined masses. The special form of the matrix (\ref{(5.8)}) allows
us to find a general method to derive the transformation $\omega$ and the mass
spectrum of the $Z$-bosons but this is too complicated to be presented here.
Anyway it is obvious that $\omega$ will depend on $\theta$, $\nu_{\hat i}$
and $\eta^{(i)}$.

Now we can appreciate the role of the parameters $\eta^{(i)}$
which determine the structure of the mass spectrum
of the gauge bosons. In the simplest  case of the homogeneous $\eta^{2}$ metric
(when $\eta^{(i)}\sim1/\sqrt{n}$) this spectrum results to be deeply
degenerated. Indeed, then the matrix (\ref{(5.8)})  will get the form
$(M^{2})_{\hat i\hat j}\sim(\delta_{\hat i\hat j}+
\nu_{\hat i}\nu_{\hat j}\tan^{2}\theta)/2n$, and, consequently, it will have
only two distinct eigenvalues:
$\sim 1/2n\cos^{2}\theta$ (for the eigenvector along the $\nu$
direction) and $\sim 1/2n$ (for all the other orthogonal
eigenvectors). Thereby we shall obtain one $Z$-boson of the mass
$\sim 1/\sqrt{2n}\cos\theta$ while the masses of all the other gauge bosons
including the $c$-number ones will be  $\sim 1/\sqrt{2n}$. This
reproduces the well known result  (i.e. $m_{ W}/m_{ Z}=
\cos\theta_{W}$) of the SM which has a Higgs mechanism equivalent
to the mHm. with the  metric (\ref{(4.17a)})
and  $\eta_{0}=0$. However, for the gauge models with $n>2$, the two values
mass spectrum corresponding to the homogeneous metric it is less likely to be
satisfactory. Thus the conclusion is that, in general, the parameters
$\eta^{(i)}$ must be chosen according to (\ref{(4.3)}) in a manner that permits
us to differentiate the masses (\ref{(5.22)}) among themselves.

\subsection{The generalized Weinberg transformation and the
coupling coefficients}
\

The whole transformation which brings the original gauge fields, $A_{\mu}^{0}$
and $A_{\mu}^{\hat i}$ into the physical ones, $A_{\mu}^{em}$ and
$Z_{\mu}^{\hat i}$, will be called the generalized Weinberg transformation
(gWt.). According to (\ref{(5.1)}), (\ref{(5.2)}) and (\ref{(5.11)}) this is:
\begin{eqnarray}\label{(5.12)}
A_{\mu}^{0}&=&A_{\mu}^{em}\cos\theta-\nu_{\hat i}
\omega^{\hat i\;\cdot}_{\cdot\;\hat j}Z_{\mu}^{\hat j}\sin\theta\nonumber\\
A_{\mu}^{\hat k}&=&\nu^{\hat k}A_{\mu}^{em}\sin\theta
+\left(\delta^{\hat k}_{\hat i}-
\nu^{\hat k}\nu_{\hat i}(1-\cos\theta)\right)
\omega^{\hat i\;\cdot}_{\cdot\;\hat j}Z_{\mu}^{\hat j}
\end{eqnarray}

Let us now turn to the spinor sector looking for the coupling coefficients
of the spinor-gauge field interactions. It is convenient to take the electric
and neutral charges in units of the elementary electric charge, $e_0$, by using
$g_{0}=g/e_0$ instead of $g$. First we shall introduce
(\ref{(5.12)}) in the interaction term of the Ld. (\ref{(3.8)}). Thus, we find
that the spinor multiplet $L^{\rho}$ (of the irep. $\rho$) has the following
electric charge matrix:
\begin{equation}\label{(5.13)}
Q^{\rho}=g_{0}\left[(D^{\rho}\cdot\nu)\sin\theta+y_{\rho}\cos\theta\right]
\end{equation}
and the $n-1$ neutral charge matrices:
\begin{equation}\label{(5.14)}
Q^{\rho}(Z^{\hat i})=
g_{0}\left[D^{\rho}_{\hat k}-\nu_{\hat k}(D^{\rho}\cdot\nu)(1-\cos\theta)
-y_{\rho}\nu_{\hat k}\sin\theta\right]\omega^{\hat k\;\cdot}_{\cdot\;\hat i}
\end{equation}
corresponding to the $n-1$ gauge fields, $Z_{\mu}^{\hat i}$. All the other
gauge fields, $A_{j\mu}^{i}$, will have the same coupling constant,
$g/\sqrt{2}$. From (\ref{(5.13)}) and (\ref{(5.14)}) we can calculate the
coupling coefficients corresponding to the fundamental
irep., $({\bf n},0)$. These are given by the electric charge matrix,
\begin{equation}\label{(5.15)}
Q\equiv diag(q_{1},q_{2},\cdots,q_{n})=g_{0}(D\cdot\nu)\sin\theta
\end{equation}
and by the neutral charge matrices,
\begin{equation}\label{(5.16)}
Q(Z^{\hat i})\equiv
diag(q^{(\hat i)}_{1},q^{(\hat i)}_{2},\cdots,q^{(\hat i)}_{n})=
g_{0}\left[D_{\hat k}-\nu_{\hat k}(D\cdot\nu)(1-\cos\theta)
\right]\omega^{\hat k\;\cdot}_{\cdot\;\hat i}
\end{equation}
All these matrices are traceless since they depend only on the generators
$D_{\hat i}$.

Now we can change again the parameterization of the model going from the
parameters, $(g,\theta,\nu_{\hat i})$, to the new ones, ($\theta, q_{i}$).
To remain in the spirit of the SM, we shall keep the angle $\theta$ as the
main parameter of the model, while $g_{0}$ and $\nu_{\hat i}$ will be expressed
by the electric charges $q_i$ of the fundamental multiplet (which
represent a set of $n-1$ independent parameters since $Tr Q=\sum_i q_i=0$).
This can be done by using the formulae:
\begin{equation}\label{(5.17)}
g_{0}\nu_{\hat i}\sin\theta=2Tr(D_{\hat i}Q)\quad,\qquad
   {g_{0}}^{2}\sin^{2}\theta=2Tr(Q^{2})
\end{equation}
and the new form of the matrix (\ref{(4.2)}):
\begin{equation}\label{(5.18)}
Y=-Q \tan\theta /\sqrt{2Tr(Q^{2})}
\end{equation}
resulted from (\ref{(5.15)}), (\ref{(2.6)}) and (\ref{(5.3)}). These will help
us to calculate the matrix (\ref{(5.8)}), the transformation $\omega$ and the
gWt. (\ref{(5.12)}) in this last  parameterization.

Based on these results we can find out the charges of each tensor
component of the multiplets $L^{\rho}$. First we shall express the neutral
charges (\ref{(5.16)}) of the fundamental irep. only on $\theta$ and $q_i$ and
then, with the help of the general rule given by
(\ref{(2.10)}) and by using (\ref{(5.13)}), (\ref{(5.14)}) and (\ref{(5.17)}),
 we find  the electric charges  of the component
$(L^{\rho})_{i_{1}\cdots i_{r}}^{j_{1}\cdots j_{s}}$
\begin{equation}\label{(5.20)}
(Q^{\rho})^{j_{1}\cdots j_{s}}_{i_{1}\cdots i_{r}}=
\sum_{\alpha=1}^{r}q_{i_{\alpha}}-\sum_{\beta=1}^{s}q_{j_{\beta}}+
           y_{\rho}\sqrt{2Tr(Q^{2})}\cot\theta
\end{equation}
and the corresponding neutral charges
\begin{equation}\label{(5.21)}
\left( Q^{\rho}(Z^{\hat i}) \right)^{j_{1}\cdots j_{s}}_{i_{1}\cdots i_{r}}=
   \sum_{\alpha=1}^{r}q^{(\hat i)}_{i_{\alpha}}-
   \sum_{\beta=1}^{s}q^{(\hat i)}_{j_{\beta}}-
          2y_{\rho}Tr(D_{\hat j}Q)\omega^{\hat j\;\cdot}_{\cdot\;\hat i}
\end{equation}
The electric charge of the gauge boson $A^{j}_{i\mu}$ is $q_{i}-q_{j}$.
Moreover, we specify that charge conjugation simultaneously changes
the signs of the electric and neutral charges of the spinor components
since $(L^{\rho})^c$ is of the irep. $\rho^*$ which has the generators
(\ref{(2.11)}).

Now, the last term of (\ref{(5.20)}) will allow us to define
different kind of chiral hypercharges (and of the second coupling constants,
$g'$,) we wish to use in concrete models.
For our future developments it is convenient to define
\begin{equation}\label{(5.30)}
\hat y=y \sqrt{2Tr(Q^{2})} \cot \theta
\end{equation}
where $y$ can be $y_{\rho}$ or $y^{(i)}$. Thus, in the following we could
use the new ireps. notation,
$(R,\hat y)$,  and the matrix $\hat Y=-Q$ instead of $Y$.
We note that these definitions
would be not consistent before to introduce the angle $\theta$ (of the
transformation (\ref{(5.2)}) which separates $A^{em}_{\mu}$)

 \section{The construction of the renormalizable models}
\

Our previous results indicate that the models with mHm. have good boson
sectors and, therefore, they will be renormalizable when an adequate spinor
sector will be added. The structure of this sector is determined by the choice
of its ireps. and by the values of one of the equivalent set of
parameters we have introduced above: $(g,y^{(i)})\sim
(g,\theta,\nu_{\hat i})\sim (\theta, q_{i})$. Finally, it must have a charge
spectrum in accordance to the particle-antiparticle principle and be free of
axial anomaly. Therefore, in order to complete our method, we need to briefly
review the specific formulation of these requirements for the pl. models.

\subsection{The particle-antiparticle principle}
\

The physical meaning of the model will strongly depend on the form  of the
mass terms of the spinor Ld. which may give rise to the masses of the Dirac
fields. These are represented here by pairs of left-handed components with
equal and opposite signs electric charges, i.e. pairs of Dirac partners as
that of the the Ld. (\ref{(a.2.7)}). In our approach, the desired good Dirac
mass terms   will appear because of the condition (\ref{(4.17)}) which,
according to (\ref{(5.18)}) and (\ref{(5.20)}), restricts each mass term to be
neutral, coupling only Dirac partners which will gain the same mass. When there
are many pairs with the same electric charges, some mixings could also occur.

Hence, a good spinor mass spectrum could be obtained only if each charged
left-handed component will have a partner  of the opposite sign of charge.
If this should not happen then the physical content of the model could
be compromised by the appearance of several charged massless components.
For this reason, it is necessary to introduce the following supplementary
requirement: the electric charge spectrum of the spinor sector must
be symmetric with respect to zero. This could be considered as the specific
form of the particle-antiparticle principle for  pl. models. Only when this
is accomplished, we could introduce adequate mass terms to produce
mass for all the Dirac fields.

However, it is difficult to find a general mathematical formulation of this
principle since this would be dependent by the concrete
values of the parameters. Therefore, in
applications, we are determined to verify step by step while building the model
if the structure of the charge spectrum remains correct and if all the needed
mass terms have been introduced. To be efficient, it is recommended to start
the construction of the model directly with the parameterization
$(\theta, q_{i})$.

\subsection{The cancellation of the axial anomaly}
\

In the case of pl. models, where we have no right-handed
multiplets, the conditions of anomaly cancellation have the same simple form as
in GUT (or in the current algebra) \cite{ANOMALY}. In terms of the hermitian
generators this amounts to have:
\begin{equation}\label{(6.1)}
\sum_{\rho}Tr(T_{\alpha}^{\rho}\{T_{\beta}^{\rho},T_{\gamma}^{\rho}\})=0
\end{equation}
for all the values $\alpha,\beta,\gamma=0,1,\cdots,n^2-1$ when the sum is
tacked over all the ireps. $\rho$ of the left-handed multiplets.

To separate the invariant factors of these equations (which depend only on the
choice of the ireps.), we shall use the form of the $U(1)_{c}$ generator,
$T_{0}^{\rho}=y_{\rho}1_{\rho}$, and the well known properties of the
hermitian $SU(n)$ generators of the ireps. $R_{\rho}$, which generalize
(\ref{(a.1.7)}) and (\ref{(a.1.8)}) according to the Wigner-Eckart
theorem. In our notations these are:
\begin{eqnarray}\label{(6.2)}
Tr(T_{a}^{\rho}T_{b}^{\rho})&=&
\frac{1}{2}\alpha(R_{\rho})\delta_{ab}\nonumber\\
Tr(T_{a}^{\rho}\{T_{b}^{\rho},T_{c}^{\rho}\})&=&\frac{1}{2}
\beta(R_{\rho})d_{abc}
\end{eqnarray}
where  $d_{abc}$ is the symmetric tensor
of (\ref{(a.1.8)}). The reduced matrix elements, $\alpha(R)$ and $\beta(R)$,
have specific values for each irep. $R$. Few examples are given in the Table
\ref{tab1}.
Furthermore for the complex conjugate ireps. we have
\begin{equation}\label{(6.3)}
\alpha(R^{*})=\alpha(R)\quad,\qquad \beta(R^{*})=-\beta(R)
\end{equation}

Now, from (\ref{(6.2)}) and the evident properties,
$Tr(T_{a}^{\rho})=0$ and $Tr(1_{\rho})=n(R_{\rho})$, we find that
(\ref{(6.1)}) are equivalent to the following conditions:
\begin{eqnarray}\label{(6.4)}
\sum_{\rho}\beta(R_{\rho})&=&0\nonumber\\
\sum_{\rho}\alpha(R_{\rho})\hat y_{\rho}&=&0\nonumber\\
\sum_{\rho}n(R_{\rho})\hat y_{\rho}^{3}&=&0
\end{eqnarray}
which will assure the cancellation of the axial anomaly. Particularly, for
$n=2$ the symmetric tensor $d_{abc}$ vanishes and, consequently, in the case
of the $SU(2)\otimes U(1)_{c}$ gauge models the first of the equations
(\ref{(6.4)}) do not occur.

On the other hand, we have seen that the spinor charge spectrum must be
symmetric with respect to zero. Consequently, we can add to (\ref{(6.4)}) the
minimal necessary condition to accomplish that,
\begin{equation}\label{(6.5)}
\sum_{\rho}Tr(Q^{\rho})\sim \sum_{\rho}n(R_{\rho})\hat y_{\rho}=0
\end{equation}
obtaining, thus, the complete set of
restrictions which should select the ireps.
of the spinor sector. We note that (\ref{(6.4)}) gives, in addition,
$\sum_{\rho}Tr((Q^{\rho})^{3}) =0$, but this will not suffice to determine
the symmetry of the charge spectrum and, therefore, adequate values of the
parameters $q_{i}$ are also needed.

\subsection{The summary of the method}
\

Now we have all the elements of the the method
we intend to propose for the construction of high unitary symmetry gauge model.
This starts with the choice of the ireps. of the spinor sector according to
(\ref{(6.4)}) and (\ref{(6.5)}). It follows to introduce suitable parameters
$q_{i}$ in order to define the desired (symmetric) charge spectrum. Then, from
(\ref{(5.18)}) and (\ref{(5.17)}) the characters $y^{(i)}$ as well as their
corresponding $\hat y^{(i)}$ can be obtained. Thus  the ireps. of the Higgs
multiplets will be completely determined only by the electric charges of the
fundamental irep.. Moreover, the  versor $\nu$ of the transformation
(\ref{(5.2)}) will also result from (\ref{(5.17)}).
As we have mentioned, the angle $\theta$ of this transformation remains the
main free parameter of the model. Furthermore, we shall take  arbitrary
 $(\eta_{0}, \eta)$ which  will represent a set of $n$ independent parameters
 because of (\ref{(5.22)}). These will be involved in the structure of the
whole gauge boson mass spectrum,  giving the masses (\ref{(5.22)}) and the
neutral boson mass matrix (\ref{(5.8)}). This matrix will be
diagonalized by the transformation $\omega$ which may be calculated in each
particular case separately. The next step is to derive the gWt. which will
point out the physical neutral gauge bosons. Their coupling coefficients will
result from (\ref{(5.21)}) and (\ref{(5.22)}). The last operation is to
introduce the constants $G$, for all the $\chi$-blocks one needs. Finally, the
resulted Ld. in unitary gauge may be written in the standard Dirac form,
(by using the formulae of the Appendix B).

We note that the parameters  $(\eta_{0}, \eta)$ could be stable under
renormalization since they appear initially only in the kinetic term of the
Higgs Ld.. Therefore, they could be kept arbitrary when we shall analyse the
properties of the model. In these conditions even the effects of the radiative
corrections would be better pointed out. For this reason we believe that the
physical interpretation, including the fit of these parameters, may be done
only after we have studied the behavior of the model.

The Weinberg-Salam model in the pl. form is the simplest example which can be
solved easily by using this method, starting directly with the parameterization
$(\theta_{W}, q_{i})$. It is of dimension $N=15$ containing the well-known
left-handed doublets, $(\nu_{L},e_{L})^T$, 3$\times (u_{L},d_{L})^T$, while
the right-handed singlets will be replaced by their charge conjugated:
$(e_{R})^c$, 3$\times (u_{R})^c$ and 3$\times (d_{R})^c$. Denoting by
$T_{i}^W$ the $SU(2)$ generators, we find that the electric charges of the
fundamental multiplet, $({\bf 2},0)$, are given by the matrix $Q=T_{3}^W$ and,
consequently, $\hat Y=-Q=-T_{3}^W$. Furthermore, from (\ref{(5.17)}) we obtain
$g\sin\theta_{W}=e_{0}$. Now it is clear that the values of $\hat y_{\rho}$
giving the desired electric charges of the spinor multiplets must be (in the
above order): -1/2, 1/6 for the doublets and 1, -2/3, 1/3 for the singlets.
On the other hand, from the form of $\hat Y$ it
results that the Higgs doublets,
$\phi^{(1)}$ and
$\phi^{(2)}$, transform like $({\bf 2}, -\frac{1}{2})$ and $({\bf 2},
\frac{1}{2})$ respectively. As we have mentioned, these doublets can be
related to the standard doublet of the Weinberg-Salam model according to
(\ref{(4.16a)}). If, in addition, we shall take the metric (\ref{(4.17a)})
then our mHm. will coincide with
its known Higgs mechanism, producing the same boson
mass spectrum. Moreover, by
using (\ref{(5.21)}) and (\ref{(5.30)}) we can find
the well-known values of the neutral charges expressed in terms of $e_0$ and
$\theta_{W}$
\cite{KH}. Thus, in our approach, we can obtain directly the final results of
the model.

This example shows that the $SU(n)\otimes U(1)_c$ pl. models with mHm. we
have proposed are a natural generalization of the SM. For this reason we
shall use the above developed technique to solve the models with
$n=3$ which would contain the SM as a submodel.

\section{$SU(3)\otimes U(1)_c$ models}
\

In the following we shall study  the $SU(3)\otimes U(1)$ electroweak
pl. models with mHm.. In fact we are interested only by the models which
contain the lepton triplet $(\nu_{eL}, e_{L}, (e_{R})^{c})^{T}$ since these
have a virtue: their gauge group combines the gauge group of the  SM to that
of the  old Pauli-Pursey-Gursey (PPG) symmetry \cite{PPG}.
This is nothing else than the $SU(2)\otimes U(1)_c$ group of the {\it maximal}
global symmetry of the Dirac  Ld. (\ref{(a.2.7)}). In addition, if one gauges
the PPG group one finds that the doubly charged boson of the actual
$SU(3)\otimes U(1)$ models is due just to this gauge. This is important since
the PPG gauge submodel is anomaly-free and, consequently, its phenomenology
could be separately treated in the future investigations related to the
electromagnetic implications of the doubly charged boson. On the other hand,
we shall see that a basis of the $su(3)$ algebra defined so that its first
three generators be those of the PPG $SU(2)$ will offer certain technical
advantages.

\subsection{Three-generations $SU(3)\otimes U(1)_c$ model}
\

Let us consider the pl. model which should have: ({\small I}) the spinor sector
of the Pisano-Pleitez model \cite{PP} put in pl. form and ({\small II}) a mHm.
with arbitrary $(\eta_{0}, \eta)$, which should satisfy (\ref{(4.3)}), and
Yukawa couplings in unitary gauge. We shall try to solve this model  supposing
that, in addition: ({\small III}) its unique coupling constant, $g$, coincides
to the first one of the SM and ({\small IV}) at least one $Z$-boson should
satisfy the condition of the SM, $m_{Z}=m_{W}/\cos\theta_{W}$.

This model is an $SU(3)_{col}\otimes SU(3)\otimes U(1)_c$ gauge
model but here we shall consider only  the electroweak interactions gauged by
$SU(3)\otimes U(1)_c$. Therefore, the color indices will be omitted,
restricting ourselves to indicate
only the triplication of the quark multiplets.
For this gauge group we shall use the hermitian $SU(3)$ generators
$T_{a}=\lambda_{a}/2$ ($a=1,\cdots,8$) represented by the Gell-Mann matrices in
the basis where
\begin{equation}\label{(6.10)}
\lambda_{3}=diag(0,1,-1) \quad,\quad \lambda_{8}=diag(-2,1,1)/\sqrt{3}
\end{equation}
Hence, the first three generators, $T_{\hat a}=\lambda_{\hat a}/2$,
$\hat a=1,2,3$, are just those of the PPG $SU(2)$ group while the $SU(2)$ group
of the SM will be generated by: $T^{W}_{1}=\lambda_{6}/2$,
$T^{W}_{2}=\lambda_{7}/2$ and $T^{W}_{3}=-(\sqrt{3}\lambda_{8}+\lambda_{3})/4$.
This is an unusual basis but it is indicated  since here $D_{1}= T_{3}$ will
be proportional to $Q$. The other diagonal generator in our previous notation
is $D_{2}=T_{8}$.

It follows to introduce the left-handed spinor multiplets in the pl. form.
Thus, all the triplets will be those of Ref. \cite{PP} (up to an unitary
transformation) while the singlets will be replaced by their charge conjugated
and, therefore, all their coupling coefficients will appear with opposite
signs. The resulted spinor sector will have the lepton triplets,
\begin{equation}\label{(6.11)}
L_{l}=
{\begin{array}{|c|}
\nu_{l}\\
\l\\
\l^{c}
\end{array}}_{L}
\sim ({\bf 3},0),\qquad l=e,\mu,\tau
\end{equation}
the quark triplets ($\times$3),
\begin{equation}
{\begin{array}{|c|}
u\\
d\\
j_{1}
\end{array}}_{L}
\sim ({\bf 3},2/3), \quad
{\begin{array}{|c|}
s\\
-c\\
j_{2}
\end{array}}_{L},\quad
{\begin{array}{|c|}
b\\
-t\\
j_{3}
\end{array}}_{L}
\sim ({\bf 3^{*}},-1/3)
\end{equation}
and the quark singlets ($\times$3)

\begin{eqnarray}
(d_{R})^{c}, (s_{R})^{c}, (b_{R})^{c}\sim ({\bf 1}, 1/3)\qquad\quad
&(u_{R})^{c}&, (c_{R})^{c}, (t_{R})^{c}\sim ({\bf 1}, -2/3)\\
(j_{1R})^{c}\sim ({\bf 1}, -5/3) \qquad\quad &(j_{2R})^{c}&,
(j_{3R})^{c}\sim ({\bf 1},4/3)\nonumber
\end{eqnarray}

The electric charge matrix of the fundamental irep., $({\bf 3},0)$, is
$Q=-2T_{3}$ while those of the ireps. $\rho=({\bf n}_{\rho}, \hat y_{\rho})$
of the above lists are given by
\begin{equation}\label{(6.12)}
Q^{\rho}=-2T^{\rho}_{3}+\hat y_{\rho}
\end{equation}
Furthermore, the condition ({\small III}) gives: $g\sin\theta_{W}=e_{0}$.
 Then, according to (\ref{(5.17)}), we obtain
\begin{equation}\label{(6.13)}
\sin\theta=2\sin\theta_{W}
\end{equation}
which fixes the value of $\theta$.  Hereby it results that
$\sin^{2}\theta_{W}\le 1/4$ \cite{NG}.
It remains to find the ireps. of the three Higgs triplets
of our mHm., $\phi^{(1)}$, $\phi^{(2)}$ and $\phi^{(3)}$. These are defined by
$\hat Y=-Q=2T_{3}$ to be,  $({\bf 3}, 0)$, $({\bf 3}, 1)$ and  $({\bf 3}, -1)$
respectively, like in Ref.\cite{PP}.

In our notations the gauge fields are $A_{\mu}^0$ and $A_{\mu}\in su(3)$
given by
\begin{equation}\label{(6.14)}
A_{\mu}=\frac{1}{2}
\begin{array}{|ccc|}
-2A_{\mu}^{8}/\sqrt{3}&\sqrt{2}W_{\mu}&\sqrt{2}V_{\mu}\\
\sqrt{2}W_{\mu}^*&A_{\mu}^{3}+A_{\mu}^{8}/\sqrt{3}&\sqrt{2}U_{\mu}\\
\sqrt{2}V_{\mu}^*&\sqrt{2}U_{\mu}^*&-A_{\mu}^{3}+A_{\mu}^{8}/\sqrt{3}\\
\end{array}
\end{equation}
where $W$ is the Weinberg charged boson while $U$ and $V$ are the new
charged bosons due to this gauge symmetry. Now one can see that the
doubly charged boson, $U$, corresponds to the non-diagonal generators of
the PPG $SU(2)$ group. It is interesting to note that it couples lepton
Cooper-like currents of the form
$g(\overline l\gamma^{\mu}(1-\gamma^{5})l^{c})/2\sqrt{2}$.

\subsection{The solution of the model}
\

The masses of the charged bosons are given by (\ref{(5.22)}), having the same
form as those of Ref. \cite{PP}. In order to find the physical neutral bosons
one  needs to calculate the transformation $\omega$,
which should diagonalize the matrix (\ref{(5.8)}), and the gWt. (\ref{(5.12)}).
To this end, it is convenient to start with the following parameterization
\begin{equation}
\eta^{2}=(1-{\eta_{0}}^2)diag(\frac{1}{2}a -b, \frac{1}{2}a+b, 1-a)
\end{equation}
Then, the masses of the charged bosons will be
\begin{equation}
{m_{W}}^{2}=m^{2}a, \quad {m_{V}}^{2}=m^{2}(1-\frac{1}{2}a-b), \quad
{m_{U}}^{2}=m^{2}(1-\frac{1}{2}a+b)
\end{equation}
where
\begin{equation}
m^{2}=\frac{1}{4}g^{2}<\phi>^{2}(1-{\eta_{0}}^{2})=
\frac{1}{2}(m_{U}^{2}+m_{V}^{2}+m_{W}^{2})
\end{equation}
and the matrix (\ref{(5.8)}) will get the form
\begin{equation}\label{(6.15)}
M^2 = m^{2}
\begin{array}{|cc|}
\frac{\textstyle 1}{\textstyle \cos^{2}\theta} (1-\frac{1}{2}a+b)&-
\frac{\textstyle 1}{\textstyle \sqrt{3} \cos\theta}
(1-\frac{3}{2}a-b))\\
-\frac{1}{\textstyle \sqrt{3} \cos\theta} (1-\frac{3}{2}a-b)&\frac{1}{3}+
\frac{1}{2}a-b\\
\end{array}
\end{equation}
We  observe that this matrix has a singularity for $\sin^{2}\theta_{W}=1/4$
and, consequently, $\sin^{2}\theta_{W}$ must remain less than $1/4$.
The transformation $\omega$  reduces here to a simple rotation of the angle
$\theta'$ (with $\omega^{1}_{1}=\omega^{2}_{2} =\cos\theta'$ and
$\omega^{2}_{1}=-\omega ^{1}_{2}=\sin\theta'$). Furthermore, we find that the
condition ({\small IV}) is satisfied if and only if
\begin{equation}\label{(6.16)}
b=-\frac{3}{2}a \tan^{2}\theta_{W}
\end{equation}
This will lead to the following {\it exact} boson mass spectrum
\begin{eqnarray}\label{(6.17)}
{m_{W}}^{2} &=&{m_{Z^2}}^{2}\cos^{2}\theta_{W}=m^{2}a\nonumber\\
{m_{Z^1}}^{2}&=&\frac{m^{2}}{1-4\sin^{2}\theta_{W}}\left(
\frac{4}{3}\cos^{2}\theta_{W}-
a\left( 1-(1-4\sin^{2}\theta_{W})\tan^{2}\theta_{W}\right) \right) \nonumber\\
{m_{V}}^{2} &=&m^{2}\left( 1-\frac{a}{2} \frac{1-4\sin^{2}
\theta_{W}}{\cos^{2}\theta_{W}}\right) \\
{m_{U}}^{2} &=&m^{2}\left(1-\frac{a}{2}
\frac{1+2\sin^{2}\theta_{W}}{\cos^{2}\theta_{W}}\right) \nonumber
\end{eqnarray}
depending on the arbitrary parameter $a \in (0,1]$. Therefore, $Z^{2}= Z$
is the Weinberg neutral boson while $Z^{1}=Z'$ is the new one of this model.

Now we have the surprise to find the angle $\theta'$
does not depend on $a$, when (\ref{(6.16)}) is accomplished. Its value is
given by
\begin{equation}\label{(6.18)}
\tan2\theta'=-\sqrt{3}\frac{\sqrt{1-4\sin^2\theta_{W}}}{1+2\sin^2\theta_{W}}
\end{equation}
Consequently, the gWt. will be also independent on $a$. Its  versor,
calculated from (\ref{(5.17)}), is  $\nu=(-1,0)$ and, therefore, it can be
written as,
\begin{eqnarray}\label{(6.19)}
A^{0}_{\mu}&=&A^{em}_{\mu}\cos\theta+(Z'_{\mu}\cos\theta'-Z_{\mu}
\sin\theta')\sin\theta\nonumber\\
A^{3}_{\mu}&=&-A^{em}_{\mu}\sin\theta+(Z'_{\mu}\cos\theta'-Z_{\mu}
\sin\theta')\cos\theta\\
A^{8}_{\mu}&=&Z'_{\mu}\sin\theta'+Z_{\mu}\cos\theta'\nonumber
\end{eqnarray}
where  $\theta$ and $\theta'$ are given by (\ref{(6.12)}) and (\ref{(6.18)})
respectively.
The gWt. allows us to calculate the coupling coefficients of the fundamental
multiplet,
$L_{l}$. Its neutral charges corresponding to the Weinberg boson $Z$ are just
those predicted by the SM while the matrix of those
corresponding to the other neutral boson, $Z'$, is
\begin{equation}
Q(Z')=-\frac{1}{\sqrt{3}\sin2\theta_{W}}\sqrt{1-4\sin^{2}\theta_{W}}
diag(1,1,-2)
\end{equation}
The neutral charge matrices of the others multiplets, $L^{\rho}$, can be
derived from (\ref{(5.21)}) and (\ref{(5.30)}). These are
\begin{eqnarray}
Q^{\rho}(Z)&=&-\frac{1}{\sin2\theta_{W}}\left( (1-4\sin^{2}
\theta_{W})T^{\rho}_{3} +
\sqrt{3}T^{\rho}_{8}+2\hat y_{\rho}\sin^{2}\theta_{W}\right)\nonumber\\
        &&\\
Q^{\rho}(Z')&=&\frac{\sqrt{1-4\sin^{2}\theta_{W}}}{\sin2\theta_{W}}\left(
\sqrt{3}T^{\rho}_{3}-T^{\rho}_{8}-2\sqrt{3}\hat y_{\rho}
\frac{\sin^{2}\theta_{W}}{1-4\sin^{2}\theta_{W}}\right)\nonumber
\end{eqnarray}
The present exact results are similar to
those of Ref.\cite{NG} obtained in the tree
approximation. In fact only the last term of $Q^{\rho}(Z')$ is different (
notice that  $\hat y_{\rho}$ coincides to the $X$-hypercharge used therein).
Moreover, we must specify that in our approach the coupling coefficients of
the vector and axial neutral currents of a fermion $f$ are
\begin{equation}
Q(Z,f)_{V}=\frac{1}{2}(Q(Z,f)-Q(Z,f^{c})) ,\quad Q(Z,f)_{A}=
-\frac{1}{2}(Q(Z,f)+Q(Z,f^{c}))
\end{equation}
because of the pl. form of the spinor sector.

Under mHm. the quark masses will be generated by the usual Yukawa interactions
involving only the fields $\phi^{(i)}$ like in Ref.\cite{PP} but
 to produce the lepton
masses we need to use our Yukawa couplings in unitary gauge. Therefore, we
shall introduce the 2-rank symmetric tensors,
$\chi_{l} = \phi^{-1} G_{l} (\phi^{(2)} \otimes \phi^{(3)}+
\phi^{(3)}\otimes \phi^{(2)})$, and the coupling
terms , $\overline{L_{l}}\chi_{l} L_{l}^{c} + h.c $, for all the leptons
($l = e, \mu, \tau$). These will give rise to the lepton masses while the
neutrinos remain massless. If we wish massive neutrinos we could use the
tensors $\chi'_{l} = \phi^{-1} G'_{l} \phi^{(1)} \otimes \phi^{(1)}$.

Hence, the model is completely solved without any kind of approximations. It
depends on the parameter $a$  which will determine the structure of the boson
mass spectrum. We observe that for $a \simeq 0$ the new gauge bosons
will be very massive as in Refs.\cite{PP,NG} while for $a \simeq 1$ their
masses become {\it smaller} than $m_{W}$. Our solution offers the possibility
to investigate both of these cases. On the other hand, the exact values of the
neutral charges we have derived bring nothing new concerning the
suppression of the flavour-changing neutral currents of this model
\cite{PP,NG,FCNC}.

\subsection{A possible new one-generation $SU(3)\otimes U(1)_c$ model}
\

Our mHm. eliminates  the usual restrictions due to the Yukawa interactions
and gives us the hope to find a new anomaly-free one-generation model in which
the suppression of the flavor-changing neutral currents is more natural
\cite{NFCNC}. However, this can not be done without introducing new particles.
There are many possibilities but here we shall restrict ourselves to present
only the most strange of them. We shall start with the observation that the
anomaly is canceled  when one uses the following ireps.: $({\bf 3}, 0)$,
3$\times ({\bf 3}, \hat y)$, 3$\times ({\bf 3}, -\hat y)$ and
$({\bf 6}^{*}, 0)$. Indeed, according to Table I and (\ref{(6.3)}), we have
$\beta({\bf 6}^{*})=-7$ and, consequently, the conditions (\ref{(6.4)}) and
(\ref{(6.5)}) are fulfilled for any $\hat y$.  Therefore, the
first irep. may be of $L_{e}$, defined by (\ref{(6.11)}), while the others
would involve the quarks $u$,  $d$ and only one exotic quark, namely that
of the electric charge $5/3$ (denoted now by $j$), as well as the corresponding
 anti-quarks.
The choice $\hat y = \frac{2}{3}$ will give the desired electric charges to the
quark triplets
\begin{equation}
3\times
{\begin{array}{|c|}
u\\
d\\
j
\end{array}}_{L}
\sim ({\bf 3}, 2/3),\qquad 3\times
{\begin{array}{|c|}
u^c\\
j^c\\
d^c
\end{array}}_{L}
\sim ({\bf 3}, -2/3)
\end{equation}
In addition, we find that the irep. $({\bf 6}^{*}, 0)$  will contain two
 massive leptons, $x$ and $y$,
of the electric charges  $q(x)=-2$ and $q(y)=-1$,
and their corresponding neutrinos, $\nu_{x}$ and $\nu_{y}$. This sextet is:
\begin{equation}
{\begin{array}{|ccc|}
\nu_{y}&y^{c}&y\\
y^{c}&x^{c}&\nu_{x}\\
y&\nu_{x}&x
\end{array}}_{L} \sim ({\bf 6^{*}}, 0)
\end{equation}

In this model the gauge and the Higgs sectors  will be identical to
those of the previous one. Therefore, all the results  obtained above
remain valid. However, differences will appear in the Yukawa
sector since here all the fermion masses may be produced only by tensors. The
masses of the quarks and of the
usual leptons  can be obtained with the help of the 2-th rank tensors like
$\chi$ and $\chi'$ defined above but the masses of the new leptons will be
produced by 4-th rank tensors. To give rise to the mass of $x$ we
need a tensor of the form $\phi^{-3}G ((\phi^{(2)})^{*}\otimes (\phi^{(2)})^{*}
\otimes (\phi^{(3)})^{*}\otimes (\phi^{(3)})^{*} + (2\Leftrightarrow 3))$ while
for $y$ we can use the tensor resulted from the symmetric direct product of
$((\phi^{(1)})^{*}\otimes (\phi^{(2)})^{*} + (1\Leftrightarrow 2))$ with
$((\phi^{(1)})^{*}\otimes (\phi^{(3)})^{*} + (1\Leftrightarrow 3))$. What is
important to observe here is that our method allows us to give independent
mass values to each spinor component, piece by piece. Thus we could
take the exotic quark $j$ and the leptons $x$ and $y$ to be very massive.

\section{Comments}
\

We have seen that the "geometrization" of the Higgs mechanism which leads to
the mHm. offers the possibility to derive a method which should help us to
easily analyse the models with high gauge symmetries. This has the advantage
to give general solutions to several technical problems related to the
model behavior (the breakdown of the gauge symmetry, the definition of the gWt.
and of the coupling coefficients, parameterizations, etc.). In its simplest
version, presented here, the mHm. has a metric introduced by hand which is
suitable for the study of the structure of the gauge boson mass spectrum
(as we have seen in the previous section). However, one could reproach that
these parameters have not a dynamical origin. This is not an impediment since
the presented mHm. can be replaced any time even  by a new extended mHm.
without a such a metric or  by an usual Higgs mechanism able to supply
the effects of this metric.

The extended mHm. can be defined by giving up the metric
$\eta^{2}$ and by introducing the new gauge invariant
field variables, $\sigma^{(i)}$, which should allow us to modify (\ref{(4.1)})
so that
\begin{equation}\label{(8.1)}
{\phi^{(i)}}^{+}\phi^{(j)}=\delta^{i}_{j}(\sigma^{(i)})^{2}
\end{equation}
Now, with an adequate Ld., each field $\sigma^{(i)}$ will gain its own vev.
and thus the system of vevs., $<\sigma^{(i)}>$, would play the role of the
eliminated metric.

The second option can be
realized by taking each $\eta^{(i)}\phi^{(i)}$ as an usual Higgs multiplet
(free of constraints) and by introducing the suitable Higgs field variables
corresponding to all the tensors involved in the Yukawa couplings in unitary
gauge. Of course, new
adequate terms in the Higgs Ld. may be also added. After this operation  we
shall  obtain a Higgs mechanism with new mass sources (i.e. the new tensor
Higgs multiplets) which could modify the
results obtained with the help of mHm.. Therefore, if one desires to conserve
these results one may arrange the contribution of the new sources to be very
small.

Finally we note that all these mechanisms will give only one mass scale.
However, one can imagine extended mHm. producing two mass scales. These
may have, in addition, an $n\times n$ matrix, $H$, of scalar fields,
like in GUT. To remain in the spirit of our mHms. we could suppose that
the fields $\phi^{(i)}$ are just the eigenvectors of this matrix corresponding
to the field eigenvalues, $h^{(i)}(x)$. When a such a mechanism with $n-k$
vector multiplets will be introduced in a $SU(n)\otimes U(1)_c$ gauge
model it could produce a good $k/(n-k)$ splitting and two mass scales.

\section{The $SU(n)$ generators}
\

Let us consider the group $SU(n)$ and its algebra, $su(n)$, in the
fundamental irep. The real generators, $H^{i}_{j}$, defined by
\begin{equation}\label{(a.1.1)}
(H^{i}_{j})_{k\; \cdot}^{\;\cdot \;l}=\delta^{i}_{k}\delta^{l}_{j}-
     \frac{1}{n}\delta^{i}_{j}\delta^{k}_{l}
\end{equation}
are $n^{2}-1$ linear independent traceless matrices because of the condition
$\sum_{i} H^{i}_{i}=0$. Their commutation relations are
\begin{equation}\label{(a.1.2)}
\left[H^{i}_{j},H^{k}_{l}\right]=\delta^{i}_{l}H^{k}_{j}-
                                 \delta^{k}_{j}H^{i}_{l}
\end{equation}

Starting with these matrices the hermitian generators $T_{a}=T_{a}^{+}$,
$a=1,2,\cdots,(n^{2}-1)$ can
be constructed \cite{SUN}. The diagonal ones are of
the form:
\begin{equation}\label{(a.1.3)}
T_{l^{2}-1}=\frac{1}{\sqrt{2l(l-1)}}\left( \sum_{i=1}^{l-1} H^{i}_{i}-
                   (l-1)H^{l}_{l}\right)
\end{equation}
for all $l=2,3,\cdots,n$. The non-diagonal generators
\begin{eqnarray}\label{(a.1.4)}
T_{(k^{2}-1)+2j-1}=\frac{1}{2}\left( H^{j}_{k+1}+H^{k+1}_{j}\right)\\
T_{(k^{2}-1)+2j}=-\frac{i}{2}\left( H^{j}_{k+1}-H^{k+1}_{j}\right)\nonumber
\end{eqnarray}
are numbered by $j=1,\cdots,k$ for each $k=1,2,\cdots,n-1$. They satisfy the
canonical commutation relations:
\begin{equation}\label{(a.1.5)}
\left[ T_{a},T_{b} \right]=i{c_{ab}}^c T_{c}
\end{equation}
where $a,b,c=1,2,\cdots,(n^{2}-1)$.
The structure constants, ${c_{ab}}^c=c_{abc}$, are
real and completely antisymmetric. Furthermore we have:
\begin{equation}\label{(a.1.6)}
\left\{ T_{a},T_{b}\right\}=\frac{1}{n}\delta_{ab}+d_{abc}T_{c}
\end{equation}
and the trace properties:
\begin{eqnarray}\label{(a.1.7)}
Tr(T_{a})=0\quad,\quad Tr(T_{a}T_{b})=\frac{1}{2}\delta_{ab}
\end{eqnarray}
which give
\begin{equation}\label{(a.1.8)}
Tr(T_{a}\lbrace T_{b},T_{c} \rbrace )=\frac{1}{2} d_{abc}
\end{equation}
Thereby $d_{abc}$ results to be completely symmetric.

The hermitian generators transform according to the adjoint irep. (Adj)
of $SU(n)$:
\begin{equation}\label{(a.1.9)}
U(\xi)T_{a}U^{+}(\xi)={Adj(\xi)_a}^bT_{b}
\end{equation}
defined as follows:
\begin{equation}\label{(a.1.10)}
Adj(\xi)=e^{-ig adj(\xi)}
\end{equation}
where $adj(\xi)=\xi^{a} adj(T_{a})$
is of the adjoint irep. of $su(n)$ for which
$adj{(T_{a})_{b}}^c=ic_{abc}$. The corresponding transformation law of the
real generators (\ref{(a.1.1)}) defines the adjoint irep. in the tensor basis
(for the multiplets with real components $\psi_j^i$ satisfying
$\psi_i^i=0$).

\section{Chirality and charge conjugation}
\

The Dirac Ld.
\begin{equation}\label{(a.2.1)}
{\cal L}_{D}=\frac{i}{2}(\overline\psi\tensor{\not\partial}\psi)-m
\overline\psi\psi
\end{equation}
which depends one the Dirac field $\psi$ and on its Dirac adjoint,
$\overline\psi=\psi^{+}\gamma^{0}$. It is well known that by using the left and
the right-handed chiral projections of
$\psi$ (i.e. $\psi_{L}=(1-\gamma^{5})\psi/2$ and
$\psi_{R}=(1+\gamma^{5})\psi/2$) this Ld. can be written as \cite{GENERAL,IZ}:
\begin{equation}\label{(a.2.2)}
{\cal L}_{D}=\frac{i}{2}(\overline \psi_{L}\tensor{\not\partial}\psi_{L}+
                   \overline \psi_{R}\tensor{\not\partial}\psi_{R})
               -m(\overline\psi_{L}\psi_{R}+\overline\psi_{R}\psi_{L})
\end{equation}

Furthermore, we have an important property namely that the charge conjugation,
$\psi\rightarrow\psi^{c}=C\overline\psi^{T}$, changes the chirality. Indeed,
bearing in mind that the matrix $C=i\gamma^{2}\gamma^{0}$ satisfies
$C=\overline C=-C^{+}=-C^{T}=-C^{-1}$
and $C^{-1}\gamma^{\mu}C=-\gamma^{\mu T}$,
we can verify the following relations:
\begin{equation}
(\psi_{L})^{c}=\frac{1+\gamma^{5}}{2}\psi^{c}=(\psi^{c})_{R} \quad,\quad
(\psi_{R})^{c}=\frac{1-\gamma^{5}}{2}\psi^{c}=(\psi^{c})_{L}
\end{equation}
Now if we consider that the spinor components anticommute
we find that two arbitrary Dirac fields, $\psi_{1}$ and $\psi_{2}$ fulfill:
\begin{equation}
\overline\psi_{1}\gamma^{\mu}\psi_{2}=
        -\overline\psi_{2}^{c}\gamma^{\mu}\psi_{1}^{c}\quad,\quad
   \overline\psi_{1}\psi_{2}=\overline\psi_{2}^{c}\psi_{1}^{c}
\end{equation}
from which we have:
\begin{equation}\label{(a.2.5)}
\overline \psi_{1R}\tensor{\not\partial}\psi_{2R}=
         (\overline \psi_{2}^{c})_{L}\tensor{\not\partial}(\psi_{1}^{c})_{L}
\quad,\quad
\overline\psi_{1L}\psi_{2R}=(\overline\psi_{2}^{c})_{L}(\psi_{1}^{c})_{R}
\end{equation}
These are the basic relations which allow us to bring any spinor sector in the
pl. form.

Particularly for only one Dirac field, $\psi$, we can do that if we put
$L_{1}=\psi_{L}$ and $L_{2}=(\psi_{R})^{c}$ so that $\psi=L_{1}+(L_{2})^{c}$
and the Ld. (\ref{(a.2.2)}) becomes:
\begin{equation}\label{(a.2.7)}
{\cal L}_{D}=\frac{i}{2}(\overline L_{1}\tensor{\not\partial}L_{1}+
                   \overline L_{2}\tensor{\not\partial}L_{2})
               -m(\overline L_{1}L_{2}^{c}+\overline{L_{2}^{c}} L_{1})
\end{equation}
In this approach the usual Dirac equations will be:
$i\not\partial L_{1}=mL_{2}^{c}$ and $i\not\partial L_{2}=mL_{1}^{c}$.
The special case is of the (neutral) Majorana field, $\psi_{M}=\psi_{M}^{c}$.
This can be written as $\psi_{M}=L+L^{c}$ depending on only one left-
handed component, $L$. Then, the corresponding Ld. has the form:
\begin{equation}\label{(a.2.9)}
{\cal L}_{D}=\frac{i}{2}(\overline L\tensor{\not\partial}L)
               -\frac{m}{2}(\overline L L^{c}+\overline{L^c} L)
\end{equation}
and the field equation (calculated by using  Grassmann derivatives) is:
$i\not\partial L=mL^{c}$.
For $m=0$, (\ref{(a.2.9)}) gives the familiar Ld. of the neutrino field.

\begin{table}[h]
\begin{center}
\begin{tabular}{|c|c|c|c|c|c|}
   & scalar \hspace*{3mm}& vector \hspace*{3mm}& antisymmetric
tensor \hspace*{3mm}&
symmetric tensor \hspace*{5mm}  & Adjoint \hspace*{3mm}\\
tensor            &      \\
R: & $\psi$ &$\psi_{i}$&$\psi_{ij}=-\psi_{ji}$&
$\psi_{ij}=\psi_{ij}$&$\psi_{j}^{i}$\\
\hline
$n(R)$& 1 & $n$ & $n(n-1)/2$ & $n(n+1)/2$ & $n^2-1$\\
\hline
$\alpha(R)$& 0 & 1 & $ n-2 $ & $n+2$ & $2n$\\
\hline
$\beta(R)$& 0 & 1 & $ n-4 $ & $n+4$ & $0$\\
\end{tabular}
\end{center}
\caption{Reduced matrix elements for the second-rank tensor ireps.}
\label{tab1}
\end{table}

\end{document}